\documentclass[a4paper,technote]{IEEEtran} 
\usepackage{cite}
\usepackage{graphicx}
\DeclareGraphicsExtensions{.pdf,.jpeg,.png}

\usepackage{pgfplots}

\usepackage{subfigure}

\usepackage[cmex10]{amsmath}
\usepackage{hyperref}
\hypersetup{
    bookmarks=true,         
    unicode=false,          
    pdftoolbar=true,        
    pdfmenubar=true,        
    pdffitwindow=false,     
    pdfstartview={FitH},    
    pdftitle={My title},    
    pdfauthor={Author},     
    pdfsubject={Subject},   
    pdfcreator={Creator},   
    pdfproducer={Producer}, 
    pdfkeywords={keyword1} {key2} {key3}, 
    pdfnewwindow=true,      
    colorlinks=true,       
    linkcolor=blue,          
    citecolor=red,        
    filecolor=magenta,      
    urlcolor=cyan           
}

\begin{document}

\title{An investigation into the Gustafsson limit for small planar antennas using optimisation}%
%
%

\author{Morteza~Shahpari, 
        David~V.~Thiel, 
        Andrew~Lewis 
\thanks{Manuscript received August 20, 2012; accepted on November 8, 2013.
This work is partly funded by a grant from Australian Research Council DP130102098.}
\thanks{M. Shahpari and D. Thiel are with Centre for Wireless Monitoring and Applications, School of Engineering, Griffith University, Nathan, Queensland 4111, Australia}
\thanks{A. Lewis is with the Institute for Integrated and Intelligent Systems, Griffith University, Queensland, Australia }
\thanks{DOI:  \href{http://dx.doi.org/10.1109/TAP.2013.2290794}{10.1109/TAP.2013.2290794}}}

%
%

\markboth{IEEE Transactions on Antennas and Propagation,~Vol.~X, No.~X, January~20XX}%
{Shahpari \MakeLowercase{\textit{et al.}}: An investigation into the Gustafsson limit for small planar antennas using optimisation}
%



\maketitle

\begin{abstract}
\boldmath
The  fundamental  limit  for  small  antennas provides  a  guide  to the  effectiveness  of designs.
Gustafsson   et  al,  Yaghjian  et  al,  and Mohammadpour-Aghdam et al independently deduced  a variation of the  Chu-Harrington limit for planar  antennas in different forms.  
Using a multi-parameter optimisation technique  based on the ant colony algorithm, planar, meander  dipole antenna designs were selected on the basis of lowest  resonant frequency  and  maximum radiation efficiency.
The  optimal antenna designs across  the  spectrum  from 570 to  1750 MHz occupying  an area of $56mm\times25mm$ were compared  with these limits calculated  using the polarizability tensor. 
The results were compared with Sievenpiper's  comparison  of published  planar  antenna properties.
The optimised antennas have greater than 90\% polarizability compared to the  containing conductive box in the range $0.3<ka<1.1$, 
so verifying the optimisation algorithm.
The generalized  absorption efficiency of the small meander line antennas is less than  50\%, and results are the same for both PEC and copper designs.
\end{abstract}


%

\section{Introduction}
%
%
%
%

Small antenna  theory  was introduced by Wheeler~\cite{Wheeler1947} for the first time and Chu~\cite{Chu1948} developed a field based method to find general formulas for the maximum gain $G$, the minimum  quality  factor  $Q$ and the maximum  $G/Q$  ratio. 
Since then Harrington~\cite{Harrington1958_TAP}, Collin and Rothschild~\cite{Collin1964_TAP}, and McLean~\cite{McLean1996_TAP} and many other authors  tried  to verify and re-examine  Chu's  findings. 
A holistic review on these works is available in \cite{Hansen_b_2011,Volakis_b_2010}.

Gustafsson  et al~\cite{Gustafsson2009_TAP,Gustafsson_2007_RSPA}  proposed a new hypothesis  which puts  another lower bound  on antennas   of arbitrary shape. 
The  new bound  states  the $D/Q$  ratio  (or  gain-bandwidth product)   is directly  related  to the generalized absorption  efficiency and polarizability  of the antenna obstacle. 
The overall antenna performance is directly dependent on the electrostatic  polarizability of the  antenna. 
The  new limit  is close to the  actual  $Q$ values of antennas while previous bounds sit far lower for real world antennas.

Vandenbosch~\cite{Vandenbosch_2010_TAP}  proposed  a  new method  to  calculate  $Q$ from the current distribution of the source. 
He later extended this method to a simple procedure to find the $Q$ of small antennas  by making the determinant of a matrix equation zero~\cite{Vandenbosch_2011_TAP}.
Also, an explicit relation has been proved between the actual  volume of the antenna and its associated $Q$ factor~\cite{Vandenbosch_2012_TAP}.

Yaghjian and Stuart~\cite{Yaghjian_2010_TAP} used a quasi-static  approximation to find the $Q$ factor.   
Their  $Q$ has  an  inverse  relationship  to  the  polarizability  of the  antenna (similar  to \cite{Gustafsson2009_TAP}).  
In~\cite{Mohammadpour_Aghdam_2011_EuMC}, the works of \cite{Yaghjian_2010_TAP} were simplified to find $Q$ for planar structures. 
In \cite{Chalas_2012_EuCAP,Capek_2012_TAP}, the characteristic modes were employed to compute the $Q$ factor.

Sievenpiper  et al~\cite{Sievenpiper2012_TAP} recently  studied  a large number  of antenna  designs published in  IEEE  Transactions on Antennas  and  Propagation, and found that the McLean-Chu limit is a valid bound on the antenna $Q$. 
Due to the significant role of bandwidth  and efficiency in engineering problems, they proposed using the $B\eta_r$ limit  instead of $Q$. 
For instance, Kanesan and Thiel~\cite{Kanesan2012_APS} concluded that a 28.5\% bandwidth  was required for planar  RFID antennas  when placed on objects with electrical properties  in the conductivity range $10^{-7} < \sigma < 3\times 10^{-2} S/m$  and  relative  permittivity $1<\epsilon_r<6$. 
Furthermore, the bandwidth plays a key role in the system design of a front end module in which antenna is a part  of the whole system.

The  Q limit by Chu-McLean  is simple and has been used for decades, however, it is usually far from practical for small antenna designs when $ka$ approaches zero. 
If the radiation  efficiency $\eta_r$  is less than  100\%, then  the $Q$ of the  antenna  decreases proportionally, and  $Q/\eta_r$      eliminates the dependency  of $Q$ to losses in the antenna structure.
The Gustafsson  method  is based on calculating  the polarizability tensor.  
This inherently  assumes that  the conducting  elements are perfectly conducting.   
This means that  the  efficiency calculations  by Gustafsson  will be the highest possible values, but the antenna $Q$ for the lossless cases are close to small antenna designs.

Thiel and Lewis et al~\cite{Galehdar2009_IJRFCAD,Lewis2009_CEC,Lewis_bc_2009}
employed the Ant Colony Optimization to search for the best performing antennas  for a given size in terms of radiation efficiency and  lower resonant frequency.   Based  on a  planar  meander-line antenna and  a  square  grid  pattern, a  range  of optimised  structures were obtained \cite{Galehdar2009_IJRFCAD,Weis2008_CEC} and a review on the effect of radius of the wire segment was reported  in~\cite{Galehdar2008_ISAPE,Galehdar2009_AWPL}.

In this paper, we used a set of previously optimized antennas to examine validity of the recent bounds on antennas. 
The  left hand side of the bound in \cite{Gustafsson2009_TAP} is the directivity and Q-factor which were calculated  using the MoM EM simulation  package FEKO \cite{FEKO}.
To find generalized absorption  efficiency $\check{\eta}_a$, each antenna was simulated by FEKO over the frequency range of 0.1-50GHz while the  resonant  frequency  is between  0.5-1.7GHz.  
The  polarizability  was calculated  by our own Method  of Moments  code for each antenna. 
In addition to the lossless antennas, the identical antennas with loss are reported  in this paper. 
The main results  of the paper are: 
(1) 
the polarizability  of the meander line is less than the polarizability of the containing box
(2) $\check{\eta}_a$ is less than  50\% for small meander lines, but approaches 50\% for the optimised antennas, 
(3) the values of $\check{\eta}_a$ for PEC and copper antennas are almost identical,
(4) the different limits  on the  performance  of the lossy antennas  are explored, 
(5) meander  line antennas  optimised  for efficiency and frequency approach  the theoretical  limit in~\cite{Gustafsson2009_TAP}, 
(6) the $B\eta_r$  limit is valid for the complete set of antennas  in both lossless and lossy cases.

\section{Background Theory}
The minimum quality factor $Q$ of an small omnidirectional  antenna was first derived  by Chu~\cite{Chu1948} and  later  by Collin~\cite{Collin1964_TAP} and  Mclean~\cite{McLean1996_TAP} for linear and  circular polarizated TE and TM waves, respectively:
\begin{eqnarray}
Q=\dfrac{1}{ka}+\dfrac{1}{(ka)^3}\\
Q=\dfrac{1}{ka}+\dfrac{1}{2(ka)^3}
\label{Q_Chu}
\end{eqnarray}
where $k$ and $a$ are the wavenumber and minimum radius of the circumscribing sphere  of antenna.  
Further research  by Pozar~\cite{Pozar2009EuCAP} demonstrated that  the lower $Q$ in \eqref{Q_Chu} is the  result  of simultaneous  use of TE  and TM modes and not the polarization  of the radiated  wave.

Sievenpiper et al~\cite{Sievenpiper2012_TAP} suggested that instead of calculating $Q$, the Bandwidth-Efficiency product, the following form should be used as a criteria  of antenna performance:
\begin{equation}
B\eta_r=\dfrac{1}{\sqrt{2}}\left(  \dfrac{1}{ka}+\dfrac{1}{n(ka)^3} \right) ^{-1}
\label{eq:Sievenpiper}
\end{equation}
where $\eta_r$ is the radiation efficiency in \eqref{Eq:eta_r} and $B$ is the fractional bandwidth. $n$ is determined by the  type of the propagation from the antenna \cite{Pozar2009EuCAP,Sievenpiper2012_TAP}.  
Equation \ref{eq:Sievenpiper} is called as the first and second order limits when $n$ is selected as 1 and 2 respectively.

Gustafsson et al~\cite{Gustafsson2009_TAP}  considered the  antenna in the  receiving mode and  looked to the  scattering properties  of the antenna. 
By invoking an optical theorem, if $\hat{e}$ and $\hat{k}$ are the polarization  and direction  of the wave propagation, one can obtain:
\begin{multline}
\int\limits_{0}^{\infty} \dfrac{\left( 1-\arrowvert\Gamma(k)\arrowvert^2\right) D }{k^4} dk ={} \\
 {} \dfrac{\check{\eta}_a k_0^3}{2\pi} \Bigg( \hat{e} \cdot \gamma_e \cdot \hat{e} + \left( \hat{k} \times \hat{e} \right) \cdot \gamma_m \cdot \left( \hat{k} \times \hat{e} \right) \Bigg)
 \label{D/Q_Gust1}
\end{multline}
In \eqref{D/Q_Gust1}, $\Gamma$ and $D$  are the antenna reflection coefficient and directivity respectively while $\gamma_e$ and $\gamma_m$ are the electric and magnetic polarizability of the antenna obstacle.
The generalized absorption efficiency, is defined by the ratio of integrated absorption cross section to the sum of the integrated absorption and scattered cross sections:
\begin{equation}
\check{\eta}_a = \dfrac{\int\limits_{0}^{\infty}\dfrac{\sigma_a}{k^2}dk}
{\int\limits_{0}^{\infty}\dfrac{\sigma_a+\sigma_s}{k^2}dk}
\label{eta_G_a}
\end{equation}
This can be interpreted as the ratio of the total absorbed power to the sum of total absorbed and scattered powers. 
Finally, the bound on $D/Q$ ratio can be written as:
\begin{eqnarray}
\dfrac{D}{Q} \leq  {} \dfrac{\check{\eta}_a k_0^3}{2\pi} \Bigg( \hat{e} \cdot \gamma_e \cdot \hat{e} + \left( \hat{k} \times \hat{e} \right) \cdot \gamma_m \cdot \left( \hat{k} \times \hat{e} \right) \Bigg)
\label{Eq_Gust_limit}
\end{eqnarray}
It is important to note that as long as no magnetic material is present in the structure $\gamma_m$ should be assumed zero \cite{Gustafsson2009_TAP}.

Yaghjian and Stuart \cite{Yaghjian_2010_TAP} used a quasi-static approximation to write a bound on $Q$ in terms of antenna volume $V$ and polarizability\footnote{Polarizability was denoted with $\alpha$ in \cite{Yaghjian_2010_TAP}, but we use $\gamma$ because of consistency with other recent publications} $\gamma$.
\begin{align}
Q_{Y,lb} = \frac{6\pi}{k^3 \gamma} (1-V/\gamma)
\label{eq:Q_Y,lb}
\end{align}

The limit in \cite{Yaghjian_2010_TAP} can be applied for both planar and non-planar  antennas.

As suggested by Mohammadpour Aghdam et~al~~\cite{Mohammadpour_Aghdam_2011_EuMC}, planar structures cannot approach the bounds for spherical shapes (i.e. Chu limit).
They considered the fact that  one can neglect the volume of the planar antennas, and simplified \eqref{eq:Q_Y,lb} in the following form:
\begin{align}
Q_{rect,lb} = \dfrac{9/2}{(ka)^3}\dfrac{1}{\gamma_l^{n}}
\label{eq:Q_M}
\end{align}
where $\gamma_l^{n}$ is ratio of the polarizability of the meander line normalized to the polarizability of the enclosing rectangle.
The Q from \eqref{eq:Q_M} is the lowest possible value for a non-magnetic planar  structure.
However, the bounds from \cite{Gustafsson2009_TAP} and \cite{Yaghjian_2010_TAP} can include the effect of the magnetic materials.

\section{Method}
\begin{figure}

\subfigure [576MHz \label{Fig:576 design}]
{
\includegraphics{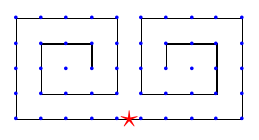}
}
\subfigure [596MHz \label{Fig:596 design}]
{
\includegraphics{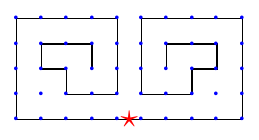}
}
\subfigure [660MHz]
{
\includegraphics{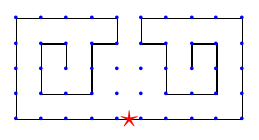}
}\\
\subfigure [724MHz\label{Fig:724 design}]
{
\includegraphics{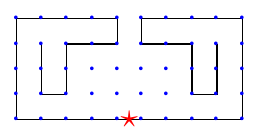}
}
\subfigure [790MHz]
{
\includegraphics{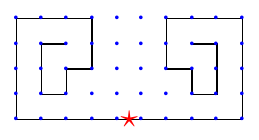}
}
\subfigure [912MHz]
{
\includegraphics{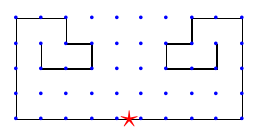}
}\\
\subfigure [1120MHz]
{
\includegraphics{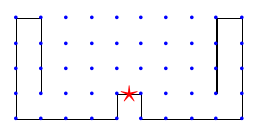}
}
\subfigure [1320MHz]
{
\includegraphics{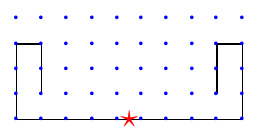}
}
\subfigure [1757MHz \label{Fig:1757 design}]
{
\includegraphics{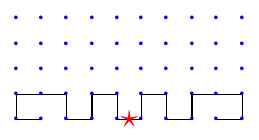}
}\\
\subfigure[1071MHz]{
\includegraphics{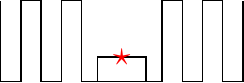}
}
\subfigure[1116MHz]{
\includegraphics{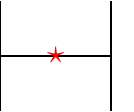}
}
\caption{Geometry of antenna and grid points (red $\star$ shows position of the feed).  (a)-(i) are optimized antennas (j) RFID tag (k) H-Shape antenna \cite{Jonsson_2009_USRI}.} 
\label{ant Geometry}
\end{figure}

The Ant Colony algorithm  by Lewis et al~\cite{Lewis2009_CEC,Lewis_bc_2009} used a rectangular matrix of points  through  which  the  conducting  meander  line passed  (see Fig.~\ref{ant Geometry}). 
The  line was terminated when the  end  of the  line has  no  further  option but  to  join  an  existing  line.   The  overall  size of the  planar  antenna   was
$56mm \times 25mm$ which gives an  aspect  ratio  of 2.24.
This is close to  the maximum performance identified by Gustafsson  et al (maximum  1.96).
The antenna optimisation   was conducted  using  NEC  as  the  solver \cite{NEC}. 
These meander line antennas are studied in \cite{Shahpari-2013-APSURSI}, in terms of their $Q$ factor and figure of merit (FOM).
After optimization, FEKO  was used to do all of the EM calculations regarding $Q$ factor, radiation  and generalized absorption efficiencies, etc using segment mesh elements.  
The  electrostatic computation of polarizability, was done using our own code which solves the  problem  using triangular mesh elements. 

Also, to make a comparison with state  of the art antennas, a RFID tag antenna, and a H-shape antenna were included in this investigation.

\newcommand{\AntShScl}{0.041}
\newcounter{mycount} 

\begin{figure}
\centering
\includegraphics{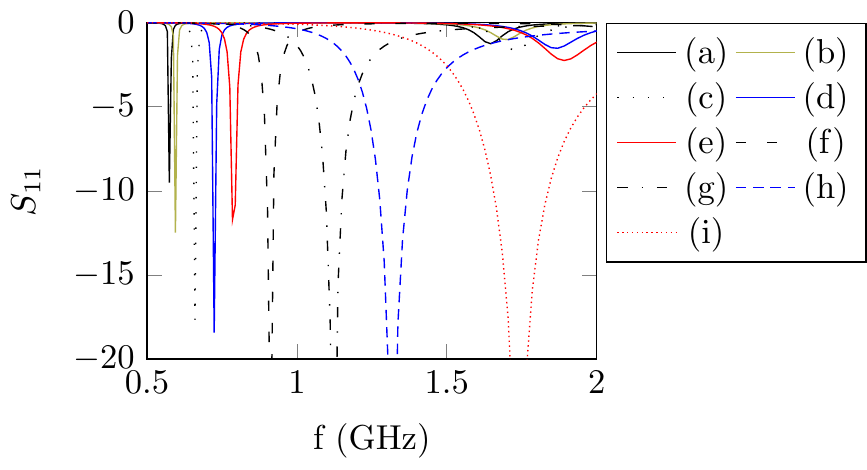}
\caption{S-parameters of the antennas in Fig.~\ref{ant Geometry}.}
\label{Fig_S11_results}
\end{figure}

In order  to  compute  Gustafsson's  limit, the  previously  optimized  antennas were analysed  using the  procedure  in \cite{Gustafsson2010_IJAP}; 
first the optimized lossy antennas  were analysed  in transmission mode to derive the impedance  and  gain using FEKO \cite{FEKO}.  
Figure~\ref{Fig_S11_results} shows the reflection coefficient of all antennas.
We  therefore calculated the Q factor  using the  Yaghijian-Best  formula \cite{Yaghjian2005_TAP} which is valid under the assumption that the impedance at the resonance is well approximated by a single mode resonance:
\begin{align}
Q=\dfrac{\omega}{2R_0(\omega)}\sqrt{[R'(\omega)]^2+[X'(\omega)+{X(\omega)/\omega}]^2},
\end{align}
where $\omega$  is resonant radian frequency, $R_0(\omega)$, and $X_0(\omega)$ are the resistance and reactance of antenna, while $R_0'(\omega)$, and $X_0'(\omega)$ are the slope of the $R_0(\omega)$, and $X_0(\omega)$.  This formulation assumes a single resonance.

The radiation efficiency was computed using the method in \cite{Galehdar2007_AWPL} which works with the summation of the loss terms in each segment of lossy wire antenna. 
When $R_{in}$ is the input impedance, $I$ is the current at the feed, $r$ is the radius of the wire, $f$ is the frequency, $\mu_0$ is the permittivity of free space, $\sigma$ is the conductivity of the conductor, $l$ is the length of the wire segment, $i_i$ is the current in the segment $i^{th}$, $\eta_r$ can be found from:
\begin{align}
\eta_r= \dfrac{R_{in} I^2 - \sqrt{\dfrac{\pi f \mu_0}{\sigma}} \dfrac{l}{2\pi r} \sum\limits_{i=1}^{N} i_i^2}{R_{in} I^2}.
\label{Eq:eta_r}
\end{align}

The antenna was centrally loaded by the impedance of the antenna at the first resonance. A co-polarized plane wave in the boresight of the antenna was used as the excitation in FEKO to find absorbed and scattered cross sections by the antenna. 

The polarizability calculation can be started from the following integral form of Laplace's equation
\begin{align}
x_{j}+C_{j} = \iint\limits_{\partial S}^{} \frac{\rho_j({x}')}{4\pi \left | x-{x}' \right |} d{S'},
\label{eq:laplace}
\end{align}
where $\rho_{j}$ is the surface charge density when the object is located in a static field of the unit amplitude in the $\hat{x}_j $ direction and integration is over the surface of the geometry $\partial S $. $x$ and $x'$ refer to observation and source points, respectively. $C_{j}$ is selected in such a way that the total charge on the object is zero:
\begin{align}
\iint\limits_{\partial S} \rho_j dS' =0.
\end{align} 

We used MoM with pulse basis functions to solve \eqref{eq:laplace}. After finding $\rho_j$ over $\partial S$, the polarizability in the $\hat{x}_i$ direction due to applied field in $\hat{x}_j$ is:
\begin{align}
\gamma_{ij} = \iint\limits_{\partial S}^{ } x_i \rho_j \left( x \right) dS.
\end{align}

\begin{figure}
\centering
\includegraphics{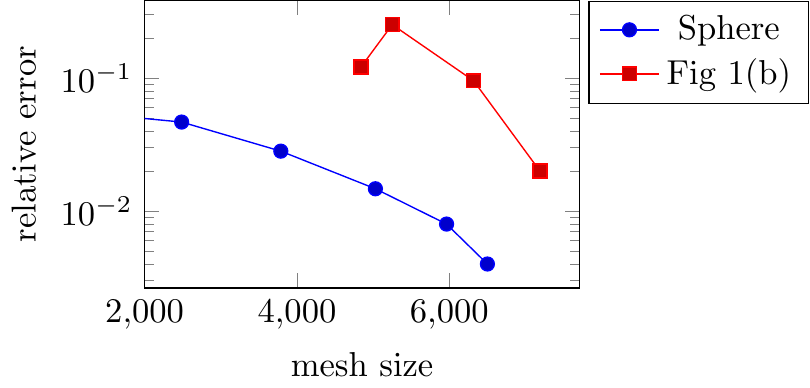}
\caption{Convergence comparison of polarizability calculations for a sphere (circle) and meander antenna design in Fig.~\ref{Fig:596 design} (square). }
\label{Fig_Convergence_Diagram}
\end{figure}

With this method \cite{Shahpari_2012_CTAT}, polarizability of each optimized antenna was computed. 
The method was verified using direct comparison with theoretical values for spheroids \cite{Shahpari_2012_CTAT}. 
Fig.~\ref{Fig_Convergence_Diagram} demonstrates that  the code shows very good convergence with an increasing number of mesh elements.
The existance of corners in the meander antennas results in less accuracy and slower congegence.

Matlab was used to compute theoretical parameters from the results of the radiation and scattering simulations. For both transmitting and receiving mode, the conductivity of the wires was assumed to be that of copper $(5.7\times 10^7 S/m)$.

It is rather  straight forward to calculate the other bounds on $Q$ \cite{Yaghjian_2010_TAP,Mohammadpour_Aghdam_2011_EuMC} once the polarizability is computed. 
It should be noted that the polarizability of each antenna should be directly substituted in \eqref{eq:Q_Y,lb}, while \eqref{eq:Q_M} needs the polarizability of antenna normalized to polarizability of the equivalent rectangle. 

\section{Results}
\subsection{Polarizablity}
\begin{figure}
\centering
\includegraphics{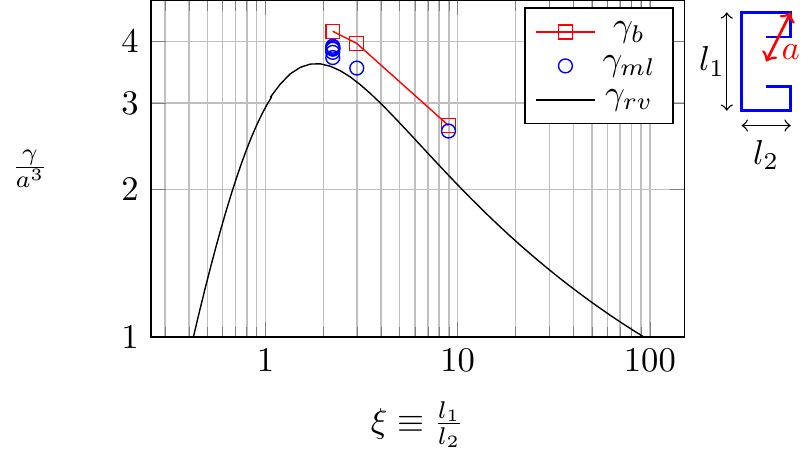}
\caption[m]{Polarizability versus length to width ratio (\begin{tikzpicture}\draw [red] (0,0)--(0.6,0); \draw [red] (0.2,-0.1) rectangle (0.4,0.1);
\end{tikzpicture} box containing antenna, 
\tikz \draw [blue,thick] (0,0) circle [radius=0.075]; meander line antenna, \tikz \draw [thick] (0,0.1)--(0.5,0.1); rectangle)}

\label{Fig_Polarizability_VS_l1_l2}
\end{figure}

Figure\ref{Fig_Polarizability_VS_l1_l2} shows the normalized polarizability of nine optimized meander line antennas $\gamma_{ml}$ as a function of the semi-axis ratio $\xi=l_1/l_2$. 
The polarizability of the designs are compared with the polarizability of the $(l_1+2r)\times (l_2+2r)$) infinitely thin rectangle $\gamma_{rv}$, and the polarizability of the ($l_1 \times l_2 \times 2r$) parallelepiped boxes $\gamma_b$ where $r$ is radius of the wire. 
For all antennas in this study, $l_1$ and $r$ were set to $56mm$,~$1mm$, but $l_2$ varies from $25mm$ to $6.25mm$.
The polarizability of the rectangle is available from \cite{Gustafsson_2011_LAPC} in the following forms for $(\xi<1)$ and $(\xi>1)$ cases respectively: 
\begin{align}
\frac{\gamma_{rv}}{a^3}   \approx \xi^2 \frac{2 \pi - 5.215 \xi - 0.108 \xi^2}{1 - 1.162 \xi + 1.712 \xi^2 - 1.222 \xi^3},\\
\frac{\gamma_{rv}}{\gamma_{sv}}  \approx \frac{1.001 + 18.098 \xi^{-1} - 11.42 \xi^{-2} + 2.266 \xi^{-3}}
{1+ 17.074 \xi^{-1}-0.309 \xi^{-2} + 24.78 \xi^{-3} } ,
\label{eq_gamma_rv_xi_more}
\end{align}
In \eqref{eq_gamma_rv_xi_more}, $\gamma_{sv}$ is the polarizability of the spheroid which is available from \cite{Gustafsson_2011_LAPC}. 
We also emphasise that polarizability of the box should be used in the calculation of the lower bound \cite{Mohammadpour_Aghdam_2011_EuMC} on the $Q$.

\begin{figure}
\centering
\includegraphics{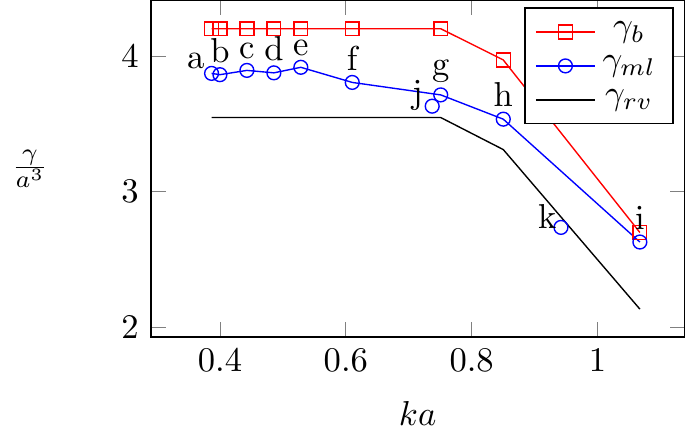}
\caption{$\gamma_{ml}$: normalized polarizability of the meander line, 
$\gamma_b$: normalized polarizability of a conducting box of dimension $l_1\times l_2 \times 2r$,
$\gamma_{rv}$:  normalized polarizability of the infinitely thin rectangle $l_1 \times l_2$.
The letter references are from Fig.~\ref{ant Geometry}. } 
\label{Fig_Polarizability_VS_ka}
\end{figure}

It is seen that the polarizability of the meander lines are less than the polarizability of the parallelepiped box which is expected from the bounds on the polarizability of the objects in \cite{Sjoberg_2009_JPhysA}. 
The reason the meander line result is higher than the rectangle result is due to the finite thickness of the wires compared to the infinitely thin rectangle.

Variation of the polarizability of the designs with the electrical length is shown in Fig.\ref{Fig_Polarizability_VS_ka}. 
It is interesting to note that the lower polarizabilities are obtained for the antennas with higher $ka$ values. 
In Fig.\ref{Fig_Polarizability_VS_l1_l2}, \ref{Fig_Polarizability_VS_ka}, the polarizabilities are normalized with respect to $a^3$ where $a$ is the radius of the circumscribing sphere of the antenna. 
For small $ka$ optimised meander lines have approximately 90\% polarizability of the box.
For $ka\geq 1$ optimised antennas have almost 99.1\% polarizability of the containing box.

\subsection{Radiation Efficiency}
\begin{figure}
\centering
\includegraphics[height=3.9cm]{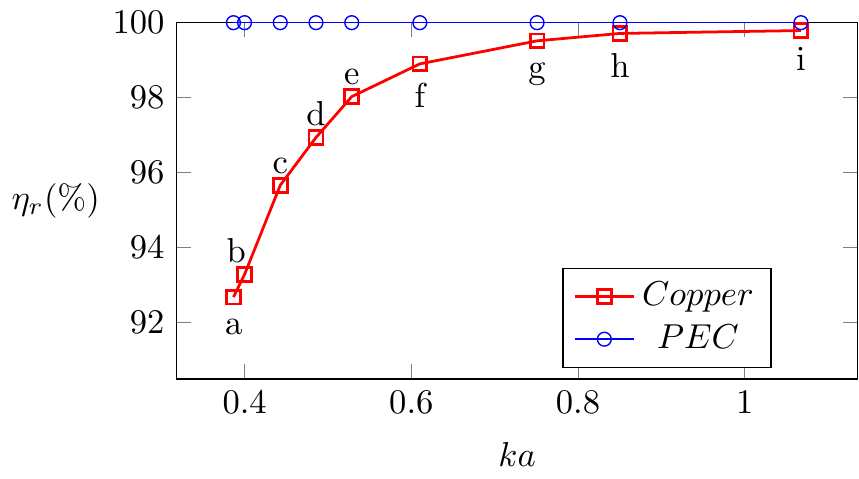}
\caption{Comparison of the radiation efficiency of the copper  and PEC antennas}
\label{Rad_Efficiency_VS_ka}
\end{figure}

Figure~\ref{Rad_Efficiency_VS_ka} shows the radiation efficiency $\eta_r$ of the meander line antennas computed from \eqref{Eq:eta_r}.  
As expected, the radiation  efficiency approaches 100\% as the  resonant frequency increases. 
This is expected as antennas with larger electrical length have a larger radiation resistance which leads to a higher $\eta_r$.

\subsection{Generalized Absorption Efficiency}
One of the parameters to be calculated is the generalized absorption efficiency $\check{\eta}_a$  (see \eqref{eta_G_a}) which can be computed by two methods:
the first method comes from the  direct  calculation  of the absorption and extinction cross sections $\sigma_a$ and $\sigma_{ext}$ . 
Therefore, simulating the antenna loaded with the resonant impedance in the receiving mode is of interest while a plane wave illuminates the antenna.
Absorbed and scattered cross sections have a direct relation with absorbed and scattered power \cite{Best2009_APM}:
\begin{eqnarray}
\sigma_a = \dfrac{1}{240 \pi} \dfrac{P_r}{\arrowvert E_i \arrowvert^2}
\label{Eq:sigm_a P_loss}
\end{eqnarray}
\begin{multline}
\sigma_{ext} = \dfrac{1}{240 \pi \arrowvert E_i \arrowvert^2} \times \\
\Bigg[P_r + P_{loss} + \int\limits_{0}^{\pi} \int\limits_{0}^{2\pi} \arrowvert E_s (\theta , \phi) \arrowvert^2   \sin(\theta) d\theta d\phi \Bigg]
\label{Eq:sigma_ext}
\end{multline}
where  $P_r$, $P_{loss}$, $E_i$ and $E_s$   are the absorbed power in the load, dissipated power in the lossy material, incident and scattered electric field strength, respectively. 
It should be noted that $P_{loss}$ has to be considered for the lossy antennas. 
The second method is to simulate the antenna in the transmitting mode and find directivity $D$ and reflection coefficient $\Gamma$. The absorption cross section was found for lossless antennas from \cite{Gustafsson2009_TAP,Gustafsson2010_IJAP}:
\begin{equation}
\sigma_a = \dfrac{\pi D(k)}{k^2} \bigg[1-\arrowvert \Gamma(k) \arrowvert^2 \bigg]
\label{Eq:sigma_a}
\end{equation}

On the other hand, the forward scattering sum rule is useful to find the overall extinct power by the antenna obstacle \cite{Gustafsson2009_TAP,Gustafsson2010_IJAP}.
\begin{equation}
\int\limits_{0}^{\infty} \dfrac{\sigma_{ext}}{k^2} dk = \dfrac{\pi}{2} \gamma_{\infty}
\label{Eq:FWD_Scat_Sum_rule}
\end{equation}
Equation \eqref{Eq:FWD_Scat_Sum_rule} can be directly used to find $\check{\eta}_a$, however, numerical integration of equations (\ref{Eq:sigm_a P_loss}-\ref{Eq:FWD_Scat_Sum_rule}) should be performed over broad range of frequencies to yield the  $\check{\eta}_a$. Both of these methods were used in calculations to ensure accuracy of the simulations.

\begin{figure}
\centering
\includegraphics[height=4cm]{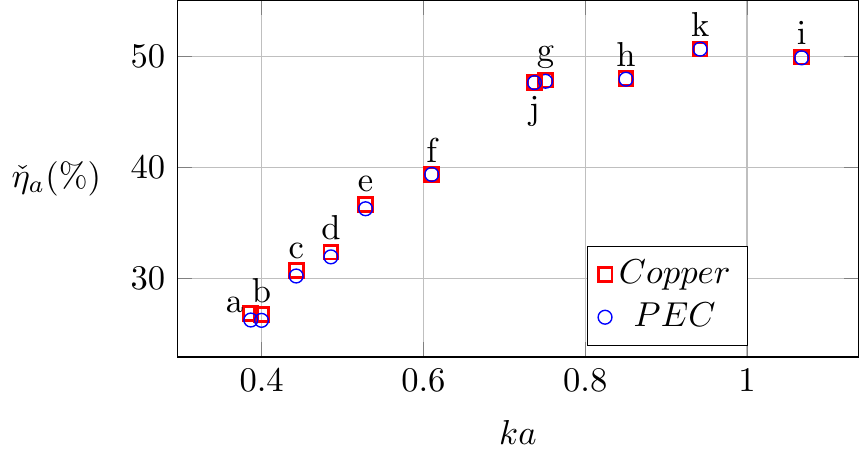}
\caption{
Explicit values $\check{\eta}_a$ for the lossy and lossless antennas while each antenna is loaded with a matched impedance at the resonant frequency.
}
\label{G_ab_Efficiency_VS_ka}
\end{figure}

Variations in $\check{\eta}_a$ with $ka$ is illustrated in the Fig.~\ref{G_ab_Efficiency_VS_ka} for both lossy and lossless antennas. 
It shows that for a small antenna $\check{\eta}_a$ is less than 50\% and approaches 50\% for antennas with larger $ka$.
Therefore, especially for electrically small antennas, one should not estimate the generalized absorption efficiency as 50\%. 
However, one should compute $\check{\eta}_a$ with the highest possible accuracy due to its crucial role in equation \eqref{Eq_Gust_limit}.

Another important result in Fig.~\ref{G_ab_Efficiency_VS_ka} is that $\check{\eta}_a$ is not sensitive to the conductivity of the material. 
Figure.~\ref{G_ab_Efficiency_VS_ka} shows that the change from PEC to copper makes  a little difference to the  $\check{\eta}_a$ values.

\begin{figure}
\centering
\includegraphics{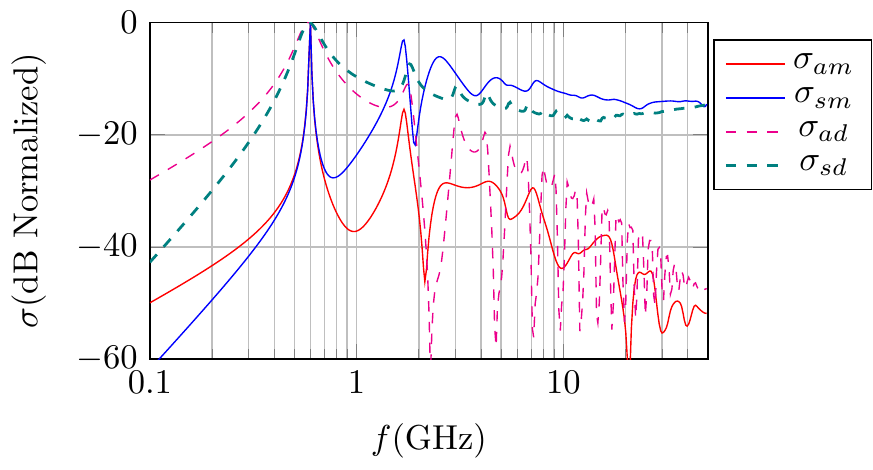}%
\caption{logarithmic plot of $\sigma_{am}$, $\sigma_{sm}$  absorption and scattering cross section of meander line (Fig.\ref{Fig:596 design}), $\sigma_{ad}$, $\sigma_{sd}$ absorption and scattering cross section of dipole}
\label{Fig_loglog_semilog}
\end{figure}

A comparison between the absorbed and scattered cross sections ($\sigma_a$ and $\sigma_s$) of the optimized meander line antennas with a classic dipole antenna provides some intuitive explanation. 
The dipole antenna is tuned to resonate at 596MHz with 244.2mm length and length to diameter ratio of 1000. Absorbed and scattered cross sections are normalized and depicted in the Fig.\ref{Fig_loglog_semilog} on logarithmic plot. 
Similar to the results in \cite{Gustafsson2009_TAP}, $\sigma_a$  and $\sigma_s$ of the dipole follow each other in the first resonance, and they have a descending magnitude envelope and bandwidth in the higher resonances. 
In the case of meander line antennas, the sharpest resonance is the first resonance, and $\sigma_s$ is much greater than the $\sigma_a$ over a broad range of frequencies. 
This means that the overall integration of the $\sigma_a$  and $\sigma_a+\sigma_s$ leads to $\check{\eta}_a\leq$50\%.

\subsection{$Q$ factor}

\begin{figure}
\centering
\includegraphics{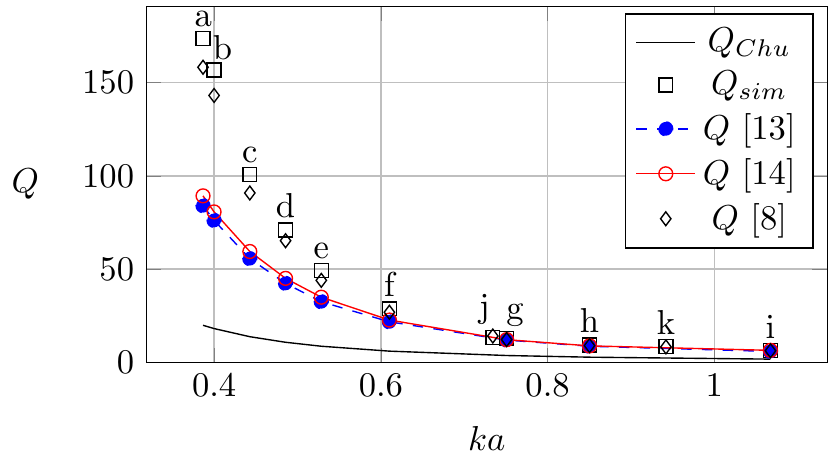}
\caption[m]{Comparison of the bounds on $Q$ by Chu, Yaghjian \cite{Yaghjian2005_TAP}, Mohammadpour\cite{Mohammadpour_Aghdam_2011_EuMC}, Gustafsson \cite{Gustafsson2009_TAP} with the practical $Q$ values.}
\label{Fig_Q_vs_ka}
\end{figure}


The $Q$ factor of the designs was compared with the limits by Chu \cite{Chu1948}, Gustafsson et~al~\cite{Gustafsson2009_TAP}, Yaghjian et~al~\cite{Yaghjian_2010_TAP} and Mohammadpour Aghdam et~al~\cite{Mohammadpour_Aghdam_2011_EuMC}. 
As is shown in Fig. \ref{Fig_Q_vs_ka}, the bound by Chu is obviously much lower than the actual $Q$ even for large values of $ka$. 
Predictions from \cite{Yaghjian_2010_TAP,Mohammadpour_Aghdam_2011_EuMC} are close together since antennas in this study do not include magnetic materials.
Limits from \cite{Yaghjian_2010_TAP, Mohammadpour_Aghdam_2011_EuMC} are closer to the practical designs for the antennas confined in $ka>0.5$. 
On the other hand, the limit in \cite{Gustafsson2009_TAP} is capable of accurately predicting $Q$ values even for small values of $ka$.
We emphasise that close predictions from Gustafsson limit happen since $\check{\eta}_a$ is rigorously calculated for each antenna separately.
\footnote{For non-magnetic planar small antennas \eqref{Eq_Gust_limit} can be reduced to \eqref{eq:Q_M} by assuming $\check{\eta}_a \approx 0.5$ and $D=3/2$.
}

The volume of the antennas decreases sequentially from Fig.~\ref{Fig:596 design} to Fig.~\ref{Fig:1757 design}. 
A general design tip proved in \cite{Vandenbosch_2012_TAP} states any increase in the actual volume, in any direction, yields a smaller $Q$ when the antenna boundaries are fixed. 
It should be emphasised that a decrease in $Q$ from a decrease in the antenna volume in this paper is not contrary to \cite{Vandenbosch_2012_TAP}. 
The implied assumption in \cite{Vandenbosch_2012_TAP} is that the frequency is unchanged. 
However, the antennas in the Fig.~\ref{ant Geometry} have different resonant frequencies.

\subsection{Bandwidth-Radiation Efficiency Product}
\begin{figure}
\centering
\includegraphics{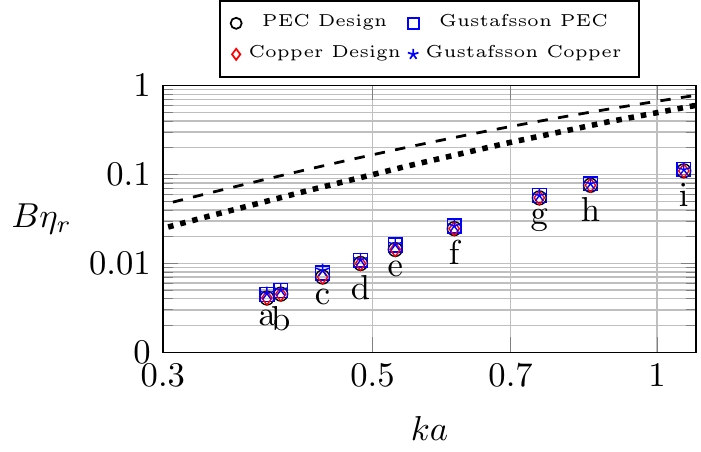}
\caption{First and second order $B\eta_r$ limit, calculated $B\eta_r$ for copper and PEC planar antennas, and $B\eta_r$  from Gustaffson's bound.}
\label{B_Eta_r_VS_ka}
\end{figure}

Figure.\ref{B_Eta_r_VS_ka} shows the $B\eta_r$  plotted as a function of $ka$ for 9 selected optimized antennas. 
Criterion for bandwidth $B$ is the range $VSWR\leq 2$ similar to the work in \cite{Sievenpiper2012_TAP}.
Included in this graph are the first and second order limits \eqref{eq:Sievenpiper} from Sievenpiper et al \cite{Sievenpiper2012_TAP}. 
The theoretical limits from Gustafsson et~al~\cite{Gustafsson2009_TAP} are also shown. 
The results for the optimised antennas show a continuous trend similar to that for the limits from Sievenpiper et al for spherical antennas. The calculation method from Gustafsson for this antennas lies slightly above the results calculated using the method explained in previous sections. 
\section{Conclusions and further work}
Fundamental limits for a set of optimized meander line antennas were analysed in this paper and compared with state of the art antennas from literature. Generalized absorption efficiency $\check{\eta}_a$  and polarizability of the antenna were computed for each of the antennas. 
The paper shows that the polarizability of the meander line antennas are bounded by the polarizability of the equivalent box. 
The computed generalized absorption efficiency illustrates that (a) $\check{\eta}_a<$50\% for small meander lines, but approaches 50\%, 
(b) $\check{\eta}_a$ is the same for PEC and Copper antennas.
Furthermore, recent limits can be reduced to the same expression for a small electric dipole antennas \cite{Gustafsson2009_TAP,Gustafsson_2007_RSPA,Yaghjian_2010_TAP,Mohammadpour_Aghdam_2011_EuMC}.
Finally, $B\eta_r$ product were studied and validated for the whole set of antennas in both lossless and lossy cases.
This paper has examined mathematical tools which allow antenna designers to compare their results with the theoretical absolute limits. The design challenge remains to find new small antenna structures which better approach the fundamental limits.

\bibliographystyle{IEEEtran}
\bibliography{IEEEabrv,TAP_ref}

\begin{thebibliography}{10}
\providecommand{\url}[1]{#1}
\csname url@samestyle\endcsname
\providecommand{\newblock}{\relax}
\providecommand{\bibinfo}[2]{#2}
\providecommand{\BIBentrySTDinterwordspacing}{\spaceskip=0pt\relax}
\providecommand{\BIBentryALTinterwordstretchfactor}{4}
\providecommand{\BIBentryALTinterwordspacing}{\spaceskip=\fontdimen2\font plus
\BIBentryALTinterwordstretchfactor\fontdimen3\font minus
  \fontdimen4\font\relax}
\providecommand{\BIBforeignlanguage}[2]{{%
\expandafter\ifx\csname l@#1\endcsname\relax
\typeout{** WARNING: IEEEtran.bst: No hyphenation pattern has been}%
\typeout{** loaded for the language `#1'. Using the pattern for}%
\typeout{** the default language instead.}%
\else
\language=\csname l@#1\endcsname
\fi
#2}}
\providecommand{\BIBdecl}{\relax}
\BIBdecl

\bibitem{Wheeler1947}
H.~Wheeler, ``Fundamental limitations of small antennas,'' \emph{Proc. {IRE}},
  vol.~35, no.~12, pp. 1479 -- 1484, Dec. 1947.

\bibitem{Chu1948}
L.~J. Chu, ``Physical limitations of omni-directional antennas,'' \emph{J.
  Appl. Phys.}, vol.~19, no.~12, pp. 1163--1175, 1948.

\bibitem{Harrington1958_TAP}
R.~Harrington, ``On the gain and beamwidth of directional antennas,'' \emph{IRE
  Transactions on Antennas and Propagation}, vol.~6, no.~3, pp. 219 --225, Jul
  1958.

\bibitem{Collin1964_TAP}
R.~Collin and S.~Rothschild, ``Evaluation of antenna {Q},'' \emph{{IEEE} Trans.
  Antennas Propag.}, vol.~12, no.~1, pp. 23 -- 27, Jan 1964.

\bibitem{McLean1996_TAP}
J.~S. McLean, ``A re-examination of the fundamental limits on the radiation {Q}
  of electrically small antennas,'' \emph{{IEEE} Trans. Antennas Propag.},
  vol.~44, no.~5, p. 672, May 1996.

\bibitem{Hansen_b_2011}
R.~C. Hansen and R.~E. Collin, \emph{Small Antenna Handbook}.\hskip 1em plus
  0.5em minus 0.4em\relax John Wiley \& sons-IEEE press, 2011.

\bibitem{Volakis_b_2010}
J.~L. Volakis, C.~C. Chen, and K.~Fujimoto, \emph{Small Antennas
  Miniaturization Techniques and Applications}.\hskip 1em plus 0.5em minus
  0.4em\relax McGraw-Hill, 2010.

\bibitem{Gustafsson2009_TAP}
M.~Gustafsson, C.~Sohl, and G.~Kristensson, ``Illustrations of new physical
  bounds on linearly polarized antennas,'' \emph{{IEEE} Trans. Antennas
  Propag.}, vol.~57, no.~5, pp. 1319 --1327, May 2009.

\bibitem{Gustafsson_2007_RSPA}
------, ``Physical limitations on antennas of arbitrary shape,'' \emph{Proc. R.
  Soc. A}, vol. 463, no. 2086, pp. 2589--2607, 2007.

\bibitem{Vandenbosch_2010_TAP}
G.~Vandenbosch, ``Reactive energies, impedance, and factor of radiating
  structures,'' \emph{{IEEE} Trans. Antennas Propag.}, vol.~58, no.~4, pp. 1112
  --1127, April 2010.

\bibitem{Vandenbosch_2011_TAP}
------, ``Simple procedure to derive lower bounds for radiation of electrically
  small devices of arbitrary topology,'' \emph{{IEEE} Trans. Antennas Propag.},
  vol.~59, no.~6, pp. 2217 --2225, June 2011.

\bibitem{Vandenbosch_2012_TAP}
------, ``Explicit relation between volume and lower bound for {Q} for small
  dipole topologies,'' \emph{{IEEE} Trans. Antennas Propag.}, vol.~60, no.~2,
  pp. 1147 --1152, Feb. 2012.

\bibitem{Yaghjian_2010_TAP}
A.~Yaghjian and H.~Stuart, ``Lower bounds on the {Q} of electrically small
  dipole antennas,'' \emph{{IEEE} Trans. Antennas Propag.}, vol.~58, no.~10,
  pp. 3114 --3121, Oct. 2010.

\bibitem{Mohammadpour_Aghdam_2011_EuMC}
K.~Mohammadpour-Aghdam, R.~Faraji-Dana, G.~Vandenbosch, S.~Radiom, and
  G.~Gielen, ``Physical bound on {Q} factor for planar antennas,'' in
  \emph{Eur. Microw. Conf.}, Oct. 2011.

\bibitem{Chalas_2012_EuCAP}
J.~Chalas, K.~Sertel, and J.~Volakis, ``Q limits for arbitrary shape antennas
  using characteristic modes,'' in \emph{{IEEE EuCAP},}, march 2012.

\bibitem{Capek_2012_TAP}
M.~Capek, P.~Hazdra, and J.~Eichler, ``A method for the evaluation of radiation
  q based on modal approach,'' \emph{{IEEE} Trans. Antennas Propag.}, vol.~60,
  no.~10, pp. 4556-- 4567, Dec. 2012.

\bibitem{Sievenpiper2012_TAP}
D.~Sievenpiper, D.~Dawson, M.~Jacob, T.~Kanar, S.~Kim, J.~Long, and
  R.~Quarfoth, ``Experimental validation of performance limits and design
  guidelines for small antennas,'' \emph{{IEEE} Trans. Antennas Propag.},
  vol.~60, no.~1, pp. 8 --19, Jan. 2012.

\bibitem{Kanesan2012_APS}
M.~Kanesan, D.~V. Thiel, and S.~G. O'Keefe, ``The effect of lossy dielectric
  objects on a {UHF} {RFID} meander line antenna,'' in \emph{IEEE Antennas
  Prop. Symp.}, Jul 2012.

\bibitem{Galehdar2009_IJRFCAD}
A.~Galehdar, D.~V. Thiel, A.~Lewis, and M.~Randall, ``Multiobjective
  optimization for small meander wire dipole antennas in a fixed area using ant
  colony system,'' \emph{Int. J. RF Microw. Comput-Aid. Eng.}, vol.~19, no.~5,
  pp. 592--597, 2009.

\bibitem{Lewis2009_CEC}
A.~Lewis, G.~Weis, M.~Randall, A.~Galehdar, and D.~Thiel, ``Optimising
  efficiency and gain of small meander line {RFID} antennas using ant colony
  system,'' in \emph{IEEE Congress on Evolutionary Computation, 2009}, May
  2009.

\bibitem{Lewis_bc_2009}
A.~Lewis, M.~Randall, A.~Galehdar, D.~Thiel, and G.~Weis, ``Using ant colony
  optimisation to construct meander-line {RFID} antennas,'' in
  \emph{Biologically-Inspired Optimisation Methods}, ser. Studies in
  Computational Intelligence, A.~Lewis, S.~Mostaghim, and M.~Randall,
  Eds.\hskip 1em plus 0.5em minus 0.4em\relax Springer Berlin Heidelberg, 2009,
  vol. 210, pp. 189--217.

\bibitem{Weis2008_CEC}
G.~Weis, A.~Lewis, M.~Randall, A.~Galehdar, and D.~Thiel, ``Local search for
  ant colony system to improve the efficiency of small meander line {RFID}
  antennas,'' in \emph{2008 IEEE Congress on Evolutionary Computation (CEC
  2008)}, 2008.

\bibitem{Galehdar2008_ISAPE}
A.~Galehdar, D.~Thiel, and S.~O'Keefe, ``Tapered wire antenna design for
  maximum efficiency and minimal environmental impact,'' in \emph{Proc. IEEE
  ISAPE}, 2008, pp. 23--26.

\bibitem{Galehdar2009_AWPL}
------, ``Tapered meander line antenna for maximum efficiency and minimal
  environmental impact,'' \emph{{IEEE} Antennas Wireless Propag. Lett.},
  vol.~8, pp. 244--247, 2009.

\bibitem{FEKO}
\emph{FEKO 6.1.1}, EM Software and Systems, 1998-2011.

\bibitem{Pozar2009EuCAP}
D.~Pozar, ``New results for minimum {Q}, maximum gain, and polarization
  properties of electrically small arbitrary antennas,'' in \emph{{IEEE
  EuCAP}}, 2009.

\bibitem{NEC}
G.~J. Burke and A.~J. Poggio, \emph{Numerical Electromagnetics Code (NEC)}, :
  National Technical Information Service (U.S. Department of Commerce), 1981.

\bibitem{Shahpari-2013-APSURSI}
M.~Shahpari, D.~V. Thiel, and A.~Lewis, ``Exploring the fundamental limits of
  planar antennas using optimization techniques,'' in \emph{{IEEE} Antennas
  Propagat. Soc. Symp. Dig.}, 2013, pp. 764--765.

\bibitem{Jonsson_2009_USRI}
B.~L.~G. Jonsson and M.~Gustafsson, ``Limitations on the effective area and
  bandwidth product for array antennas,'' in \emph{URSI Int. Symp.
  Electromagnetic Theory (EMTS)}, 2010, pp. 711--714.

\bibitem{Gustafsson2010_IJAP}
M.~Gustafsson, M.~Cismasu, and S.~Nordebo, ``Absorption efficiency and physical
  bounds on antennas,'' \emph{Int. J. Antennas Propag.}, vol. 2010, 2010.

\bibitem{Yaghjian2005_TAP}
A.~Yaghjian and S.~Best, ``Impedance, bandwidth, and {Q} of antennas,''
  \emph{{IEEE} Trans. Antennas Propag.}, vol.~53, no.~4, pp. 1298--1324, 2005.

\bibitem{Galehdar2007_AWPL}
A.~Galehdar, D.~Thiel, and S.~O'Keefe, ``Antenna efficiency calculations for
  electrically small {RFID} antennas,'' \emph{{IEEE} Antennas Wireless Propag.
  Lett.}, vol.~6, pp. 156--159, 2007.

\bibitem{Shahpari_2012_CTAT}
\BIBentryALTinterwordspacing
M.~Shahpari, D.~V. Thiel, and A.~Lewis, ``Polarizablity of 2d and 3d conducting
  objects using method of moments,'' \emph{ANZIAM Journal}, vol.~54, pp.
  C446--C458, 2013. [Online]. Available:
  \url{http://journal.austms.org.au/ojs/index.php/ANZIAMJ/article/view/6405}
\BIBentrySTDinterwordspacing

\bibitem{Gustafsson_2011_LAPC}
M.~Gustafsson, ``Physical bounds on antennas of arbitrary shape,'' in
  \emph{Loughborough Antennas and Propagation Conference (LAPC)}, Nov. 2011.

\bibitem{Sjoberg_2009_JPhysA}
D.~Sj{\"o}berg, ``Variational principles for the static electric and magnetic
  polarizabilities of anisotropic media with perfect electric conductor
  inclusions,'' \emph{J. Phys. A-Math. Theor.}, vol.~42, p. 335403, 2009.

\bibitem{Best2009_APM}
S.~Best and B.~Kaanta, ``A tutorial on the receiving and scattering properties
  of antennas,'' \emph{{IEEE} Antennas Propag. Mag.}, vol.~51, no.~5, pp. 26
  --37, oct. 2009.

\end{thebibliography}
%

%




\end{document}